\documentclass[12pt]{article}

\usepackage{amsmath,amssymb,latexsym,theorem,epsfig,amscd}
\usepackage[all]{xy}
\CompileMatrices

\newcommand{\be}{\begin{equation}}
\newcommand{\ee}{\end{equation}}

\newcommand{\rk}{{\text{ rk} }}

\newcommand{\tr}{{\text{tr}\,}}
\def\bl{\bigl(}
\def\br{\bigr)}
\def\sta{^{\ast}}

\newcommand{\les}[8]{\xymatrix{       &      & ...  \ar[r]  &  {#1}    \ar@{->} `r[d] `[l] `^dl[dlll]  `^dr/14pt[dll]    [dll] \\
&  {#2} \ar[r] & {#3} \ar[r] & {#4}  \ar `r/10pt[d] `[l]  `^dl[dlll]  `^dr/14pt[dll]   [dll] \\ 
& {#5} \ar[r]  & {#6} \ar[r] & {#7}  \ar `r/10pt[d] `[l]  `^dl[dlll]  `^dr/14pt[dll]   [dll] \\
&  {#8} \ar[r] & ... & & }}
\newcommand{\rs}[5]{
\xymatrix{
         &            &            &  0         & \\ 
         &            &            & {#5} \ar[u] & \\
0 \ar[r] & {#1} \ar[r] & {#2} \ar[r] & {#3} \ar[r] \ar[u] & 0 \\  
         &            &            & {#4} \ar[u]\ar@{.>}[ul] & \\
         &            &            &  0  \ar[u] &
}}

\newcommand{\ts}[4]
{\xymatrix{
0 \ar[r] & {#1} \ar[r] & {#2} \ar[r] & {#3} \ar[r] & 0 \\  
         &            & {#4} \ar[u] \ar@{.>}[ul]^{\gamma} \ar@{.>}[ur]_{\lambda}  &            & \\
         &            &  0  \ar[u] &            &
}}

{\theorembodyfont{\rmfamily} }
{\theorembodyfont{\rmfamily} }

\numberwithin{equation}{section}

\begin{document}

\begin{titlepage}

\vspace{-5cm}
\title{
   \hfill{\normalsize hep-th/0605203} \\[1em] {\LARGE Yukawa Couplings on Quintic Threefolds}
\\
[1em] }
\author{
     Ron Donagi$^1$, Ren\'e Reinbacher$^2$ and Shing-Tung-Yau$^3$\\[0.5em] {\normalsize
     $^1$Department of Mathematics, University of Pennsylvania}
     \\[-0.4em] {\normalsize Philadelphia, PA 19104-6395}\\[0.5em] {\normalsize
     $^2$Department of Physics and Astronomy, Rutgers University}
     \\[-0.4em] {\normalsize Piscataway, NJ 08854}\\[0.5em] {\normalsize
     $^3$Department of Mathematics, Harvard University}
     \\[-0.4em] {\normalsize Boston, MA 02138}}

\date{}
\maketitle
\begin{abstract}
\noindent
We compute the particle spectrum and some of the Yukawa couplings for a
family of heterotic compactifications on quintic threefolds $X$ involving
bundles that are deformations of $TX\oplus \mathcal{O}_X$. These are then related to the
compactifications with torsion found recently by Li and Yau. We compute
the spectrum and the Yukawa couplings for generic bundles on generic
quintics, as well as for certain stable non-generic bundles on the special
Dwork quintics. In all our computations we keep the dependence on the vector bundle moduli explicit.
We also show that on any smooth quintic there
exists a deformation of the bundle $TX\oplus \mathcal{O}_X$ whose Kodaira-Spencer class obeys
the Li-Yau non-degeneracy conditions and admits a non-vanishing triple
pairing.
\end{abstract}

\thispagestyle{empty}

\end{titlepage}

\section{Introduction}
In their proposed superstring compactification  \cite{CHSW}, Candelas, Horowitz, Strominger and Witten 
took the product of a maximal symmetric four dimensional space-time with a six dimensional Calabi-Yau 
manifold $X$ as the ten dimensional space-time. In addition, they identified the Yang-Mills connection 
with the $SU(3)$ connection of the Calabi-Yau metric and set the dilaton to be a constant. In conformal field theory language these compactifications are referred to as $(2,2)$ models since they admit a $(2,2)$ world sheet supersymmetry. Using explicit 
formulas for the four dimensional super and Kaehler potentials \cite{SW,S}, models with three chiral 
families have been studied in some depth \cite{DGKM,CK,DG,GKMR}. 
In these models, the breaking of the $E_6$ gauge group  to a $GUT$  group  or  to the Standard Model gauge group was done at the field theoretic level. 

A proposal of Witten \cite{W}  was to use bundles with $SU(4)$ 
or $SU(5)$  structure group in order for the $GUT$ group to arise as the gauge group at the string level. Mathematically, this approach relies on the Donaldson-Uhlenbeck-Yau theorem 
\cite{Do,UY} about the existence of Hermite-Yang-Mills connections on stable bundles. In conformal field theory language these models are  referred to as $(0,2) $ models. The most widely 
used  technique to find such bundles was Monad construction \cite{Ma,DG}. 

During the second string 
revolution, Horava-Witten \cite{HW} proposed a string compactification which relaxed the Green-Schwarz 
Anomaly cancellation condition by allowing $M5$ branes. Using the newly gained freedom and 
a recent mathematical technique called Spectral cover construction \cite{FMW, D1}, several $GUT$ models 
with three families were found \cite{DLOW}. Using Spectral cover construction in conjunction with constructing stable bundles as non-trival extensions, a variety  
of heterotic Standard Model vacua was found \cite{DOPW,DOPR,BD}. In particular, these constructions  involve  building stable bundles on Calabi-Yau threefolds with non-trivial fundamental groups.

In \cite{S1}, Strominger analyzed a more general heterotic superstring background  by allowing non-zero torsion and  a scalar ``warp factor''  $D(x)$  in the
metric. More specifically, he considered a ten dimensional space-time that is the
product $M\times X$ of a maximally symmetric four dimensional
space-time $M$ and an internal space $X$ such that the metric $g_{MN}$ on $M\times X$
takes the form
$$g_{MN}(m,x)=e^{2D(x)}\left(%
\begin{array}{cc}
g_{\mu\nu}(m)&0\\
0&g_{ij}(x)\\
\end{array}
\right),\quad m\in M, \qquad x\in X.
$$
Strominger showed that in
order to achieve space-time supersymmetry the internal six
manifold $X$ must be a complex manifold with a non-vanishing
holomorphic three form $\Omega$ and the dilaton field $\phi$ must be identified with $D(x)$. In addition, the gauge connection on the heterotic vector bundle $E$ over $X$ has to be hermitian Yang-Mills with respect to  the hermitian form $\omega=\sqrt{-1}g_{i\bar
j}dz^i d\bar z^j$. In summary he proposed to solve the system 
\begin{eqnarray}
\label{eq-Strom}
F\wedge\omega^2&=&0\\
F^{2,0}=F^{0,2}&=&0\\
\partial\bar{\partial} \omega&=&
\sqrt{-1}tr F\wedge F-\sqrt{-1}tr R\wedge R\\
d\sta \omega&=&\sqrt{-1}(\partial-\bar{\partial})\ln\|\Omega\|.
\end{eqnarray}
Here $R$ denotes the curvature tensor of
the hermitian metric $\omega$, $F$  the curvature of the vector
bundle $E$, $\tr$ is the trace of the endomorphism bundle of
either $E$ or $TX$, and the norm $||\cdot || $ and  dualization $d^*$ in the last equation are taken with respect to $\omega$.

Given such solution, Strominger shows that the Kalb-Ramond field $H$, the dilaton $\phi$ and the physical metric $g_{ij}^0$ are given by
$$H=\frac{\sqrt{-1}}{2}(\bar{\partial}-\partial)\omega,\;\;\;
\phi=\frac{1}{8}\ln\|\Omega\|+\phi_0,\;\;\;g_{ij}^0=e^{2\phi_0}\|\Omega\|^{\frac{1}{4}}g_{ij},
$$
where  $\phi_0$ is an arbitrary constant. 

In a recent paper \cite{LY} Li and Yau have given the first irreducible non-singular solution of this system of equations for $SU(4)$ and $SU(5)$ principal bundles. In more concrete terms, they consider  a smooth Calabi-Yau threefold
$(X,\omega_0)$ and the vector bundle
\begin{equation}
\label{ }
({E},D_0^{''})=\mathcal{O}_X^{\oplus r} \oplus TX,
\end{equation}
where $D_0^{''}$ denotes the standard holomorphic structure on $\mathcal{O}_X^{\oplus r} \oplus TX$. In addition, they consider $h_1$ to be the constant metric on $\mathcal{O}_X^{\oplus r}$ and $h_2$ the  metric on the Calabi-Yau threefold $X$ induced by the Kaehler form $\omega_0$, so that $\det(h_1\oplus h_2)$ is the constant metric on $\wedge^{r+3}E=\mathcal{O}_X$. The pair $h_1\oplus h_2$ and $\omega_0$ is a reducible solution of Strominger's system with vanishing $H$. Li and Yau show that under certain algebraic conditions, small perturbations $(E,D^{''}_s)$ give irreducible solutions of Strominger's system.

More precisely, they consider a small perturbation $(E,D^{''}_s)$ 
and its first order approximation which are described by its  Kodaira-Spencer class in
\begin{equation}
\label{ }
H^1_{\bar{\partial}}(X,{E}^{*}\otimes {E}).
\end{equation}
This vector space consists of two by two matrices whose off diagonal terms correspond to a column vector $(\alpha_1,\ldots,\alpha_r)^{t}$ where $\alpha_i \in H^1(X,TX^{*})$ and a row vector $(\beta_1,\ldots,\beta_r)$ with $\beta_i \in H^1(X,TX)$. Under the assumption that for the chosen deformation $(E,D_s^{''})$, $\{\alpha_i\}$ and $\{\beta_i\}$ are separately linearly independent, Li and Yau \cite{LY} show that there exists a family of pairs of hermitian metrics and hermitian forms $(h_s,\omega_s)$ solving  Strominger's system for the holomorphic bundle $({E},D_s^{''})$. In the limit as $s$ goes to zero, this solution converges to a metric $h_0$ whose metric connection is a hermitian Yang-Mills connection of $E$ over $(X,\omega_0)$.

In this paper we consider deformations $\mathcal{E}=(E,D_s^{''})$ of the rank four vector bundle
\begin{equation}
\label{ }
E=\mathcal{O}_X\oplus TX
\end{equation}
on smooth quintic threefolds $X$. We compute the cohomology groups for $\mathcal{E}$ and $\wedge^2 \mathcal{E}$ and the triple pairing
\begin{equation}
\label{triplepair}
H^1(X,\mathcal{E})\times H^1(X,\mathcal{E})\times H^1(X,\wedge^2\mathcal{E}) \to \mathbb{C}.
\end{equation}
We will keep the dependence on the bundle moduli in our computations explicit which gives, in particular, an explicit expression for the triple pairing \ref{triplepair}\footnote{R.R thanks B. Ovrut for pointing out the importance of such parameterization.}.

In section~\ref{stableSU(4)} we give a short review of the family $\tilde{\mathcal{E}}$ of deformations of $\mathcal{O}_X\oplus TX$ whose generic member $\mathcal{E}$ obeys the condition on its Kodaira-Spencer class required by Li and Yau. Note that this condition is also sufficient to guarantee the stability of $\mathcal{E}$ with respect to the Kaehler class $\omega_0$.

In section~\ref{se-cohgen} we compute $H^{*}(X,\mathcal{E})$ and $H^{*}(X,\wedge^2 \mathcal{E})$ for a generic $\mathcal{E}$ on
a generic quintic $X$.  We find as the only non-vanishing group $h^1(X,\mathcal{E})=100$.

In section~\ref{sect-Es} we begin our analysis for non-generic compactifications. We fix $X$ to be the Dwork quintic threefold and consider a specific $\mathcal{E}$. We find non-vanishing $H^{*}(X,\mathcal{E})$ and $H^{*}(X,\wedge^2 \mathcal{E})$. Since $\mathcal{E}$ is not generic, we include an algebraic proof of stability. In section~\ref{se-Yukawa} we compute the triple pairing \ref{triplepair} for the chosen bundle and give an explicit parameterization of its moduli dependence.

In section~\ref{se-genquintic} we generalize these results. We show that for any smooth quintic there exists a deformation $\mathcal{E}$ whose Kodaira-Spencer class obeys the above stated conditions and admits non-vanishing triple pairing. 

To place the results in physical context, let us recall the general strategy of finding vacua in heterotic string theory with spacetime supersymmetry. Instead of solving the full fledged string equations of motion, which  include all massive modes, one finds a  supersymmetric configuration in the field theory approximation. In particular, one solves for a bosonic configuration in which all fermions can be consistently set to zero. Hence the supersymmetry variations for the gravitino $\psi_M$, the dilatino $\lambda$, and the gluino $\chi$
\begin{eqnarray}
\delta \psi_M&=& \nabla_M\epsilon +\frac{1}{48}e^{2\phi}\bl
\gamma_M H-12 H_M\br \epsilon\\
\delta\lambda &=& \nabla\phi
\epsilon+\frac{1}{24}e^{2\phi}H\epsilon\\
\delta \chi &=& e^{\phi} F_{ij}\Gamma^{ij}\epsilon
\end{eqnarray}
have to vanish. Here $\epsilon$ denotes the  Majorana-Weyl spinor  which generates the supersymmetry transformation in the field theory limit.
These  equations have to be supplemented by the anomaly cancellation condition
\begin{equation}
\label{ }
dH=\tr R\wedge R-\tr F\wedge F.
\end{equation}
Having found a solution to these equation, one has found, in particular, a supersymmetric solution of the string theory equations of motion to lowest order in the dimensionless parameter $\frac{\alpha'}{R^2}$ where $R^2$ denotes the radius of the compact space $X$. Using non-renomalization theorems for the effective four dimensional superpotential one can argue that such solution can be completed to a solution to any finite power of $\frac{\alpha'}{R^2}$ in a perturbative expansion. Non-perturbative corrections have to be considered separately.  One important feature of the full solution is that it will not modify existing zero modes of the four dimensional theory, that is, it does not modify the four dimensional particle spectrum. 

The simplest way to find a solution of the equations above is to set $H=d\phi=0$. An exact solution to the equations is given by the choice of a Calabi-Yau threefold $X$ with a Kaehler metric. The gauge connection is determined by the spin-connection of $X$. These compatifications are the previously mentioned $(2,2)$ models, and their massless particle spectrum, which is charged under the low energy gauge group, is determined by the cohomology of $X$.

A generalization of these solutions is to consider the equations as lowest order approximations in $\frac{\alpha'}{R^2}$ and solve them order by order. In particular, one can start again by setting $H=d\phi=0$, and solve the supersymmetry variations by choosing a Calabi-Yau threefold $X$ with Kaehler metric, but choose a general gauge connection on a vector bundle $E$ which solves the hermitian Yang-Mills equations. The anomaly cancellation condition implies that $H$ vanishes to order $\frac{1}{R}$, but will be generically non-zero to order $\frac{1}{R^3}$. Witten argued that this correction is of string theoretic nature and that these solutions can also be consistently computed to any finite order of $\frac{\alpha'}{R^2}$. These compactifications are referred to as $(0,2)$ models and their charged massless particle spectrum is determined by the cohomology of $E$ and $\wedge^2 E$.  Note that these models have already non-vanishing torsion $H$.

The 1-parameter family  $(h_s,\omega_s)$ of solutions of Strominger's
equations found by Li and Yau determines, in particular, a 1-parameter
family $(E,D^{''}_s)$ of complex structures on the underlying vector bundle. It
is plausible, but not obvious to us, that their 1-parameter family
$(h_s,\omega_s)$  can be obtained via Witten's procedure, i.e. that these
are the $\alpha'$  corrections of the $(0,2)$ models $(E,D^{''}_s)$ determined by $h_s$,
for a suitable choice of $\alpha'=\alpha'(s)$. If this is indeed so, it would
follow from the non-renormalization theorems that the particle spectrum of
the Li-Yau solutions with torsion is identical with that of the $(0,2)$
models which we analyze in this paper. Some progress towards determining
the corrected particle content of models with torsion has been made
recently in \cite{CL}.

\section{Irreducible $SU(4)$ bundles}\label{stableSU(4)}
In this section we review the explicit construction \cite{LY} of the family $\tilde{\mathcal{E}}$ of deformations of 
\begin{equation}
\label{ }
\mathcal{O}_X\oplus TX,
\end{equation}
such that for a generic member  $\mathcal{E}$ the off-diagonal terms in the Kodaira-Spencer class do not vanish. 

To begin with, consider the combination of the normal bundle sequence with the Euler sequence.
\begin{equation}
\label{ }
{\xymatrix{    
  &{\mathcal{O}}_X \ar[d]& \\
 &{\mathcal{O}}_X(1)^{\oplus 5} \ar[d]
& \\
TX \ar[r] &T{\mathbb{P}_4}|_X \ar[r]^{\phi}&{\mathcal{O}}_X(5).
}}
\end{equation}
This defines a non-trivial canonical  extension $F$
\begin{equation}
\label{Fasext}
{\xymatrix{    
{\mathcal{O}}_X \ar[r]\ar[d] &{\mathcal{O}}_X \ar[d]& \\
F \ar[r]\ar[d] &{\mathcal{O}}_X(1)^{\oplus 5} \ar[r]\ar[d]
&{\mathcal{O}}_X(5)\ar[d] \\
TX \ar[r] &T{\mathbb{P}_4}|_X \ar[r] &{\mathcal{O}}_X(5),
}}
\end{equation}
which corresponds (up to rescaling) to the  unique element $\beta$ in $H^1(X,TX^{*})$. Using $\beta$, it is straighforward to construct a family $\tilde{\mathcal{F}}$ describing a deformation of the $\mathcal{O}_X\oplus TX$ such that its Kodaira-Spencer class has the form $\left(\begin{array}{c|c}0 & 0 \\\hline \beta & 0\end{array}\right)$. Consider the two projections
\begin{equation}
\label{ }
\pi_1: X \times A^1 \to X,\;\;\;\pi_2: X \times A^1 \to A^1,
\end{equation}
and let $t$ be the standard coordinate function on $A^1$. The class 
\begin{equation}
\label{ }
t\cdot \beta \in Ext^1_{X\times A^1}(\pi_1^{*}TX,\mathcal{O}_{X\times A^1})=H^0(A^1,\mathcal{O}_{A^1})\otimes Ext^1_X(TX,\mathcal{O}_X)
\end{equation}
defines a locally free  extension $\tilde{\mathcal{F}}$ on $X\times A^1$. Its restriction  $\mathcal{F}_t$ to $X\times t$  is isomorphic to $F$ for non-vanishing $t$ and isomorphic to $\mathcal{O}_X\oplus TX$ for $t=0$.

We will now construct a deformation of $\mathcal{O}_X\oplus TX$ such that its Kodaira-Spencer class is of the form $\left(\begin{array}{c|c}0 & \alpha \\\hline 0 & 0\end{array}\right)$ with $\alpha\neq 0$. Pick a section $u \in H^0(X,\mathcal{O}_X(5))$ and define a vector bundle $\tilde{\mathcal{F}}^{'}$ on $X\times A^1$ as the kernel of  the map $\Phi=(tu,\pi^{*}_1\phi)$, that is 
\begin{equation}
\label{ }
{\xymatrix{    
{\tilde{\mathcal{F}}}^{'} \ar[r] & {\mathcal{O}}_{X\times A^{1}} \oplus \pi^{*}_1T {\mathbb{P}}_4|_X \ar[r]^-{\Phi} &\pi^{*}_1{\mathcal{O}}_X (5). }}
\end{equation}
Note that this sequence fits naturally into the diagram
\begin{equation}
\label{ }
{\xymatrix{    
\pi^{*}_1TX \ar[d]\ar[r]& \pi^{*}_1TX\ar[d]\\
{\tilde{\mathcal{F}}}^{'} \ar[r]\ar[d] & {\mathcal{O}}_{X\times A^{1}} \oplus \pi^{*}_1T {\mathbb{P}}_4|_X \ar[r]^-{\Phi}\ar[d]^{\mathbb{I}\oplus\pi_1^{*}\phi}&\pi^{*}_1{\mathcal{O}}_X (5)\ar[d]\\
{\mathcal{O}}_{X\times A^{1}} \ar[r] & {\mathcal{O}}_{X\times A^{1}} \oplus \pi^{*}_1{\mathcal{O}}_X(5) \ar[r] &\pi^{*}_1{\mathcal{O}}_X(5)
}}
\end{equation}
Denote the restriction of $\tilde{\mathcal{F}}^{'}$ to $X\times t$ by $\mathcal{F}^{'}_t$. Clearly, $\mathcal{F}^{'}_0=\mathcal{O}_X\oplus TX$ and the Kodaira-Spencer class is for a generic $t$ and $u$ is   $\left(\begin{array}{c|c}0 & \alpha \\\hline 0 & 0\end{array}\right)$ \cite{LY}.

Finally, we will construct a family of holomorphic bundles which contains  $\tilde{\mathcal{F}}$ and $\tilde{\mathcal{F}}^{'}$ as subfamilies, that is it has to contain
\begin{equation}
\label{ }
\mathcal{F}_t \to \mathcal{O}_X(1)^{\oplus 5} \to \mathcal{O}_X(5),\;\;\;\mathcal{F}^{'}_t \to \mathcal{O}_X\oplus T\mathbb{P}_4|_X,  \to \mathcal{O}_X(5)
\end{equation}
with
\begin{equation}
\label{ }
\mathcal{F}_0=\mathcal{F}^{'}_0=\mathcal{O}_X\oplus TX.
\end{equation}
The required family will be given by a universal bundle over the total space of a vector bundle. The base $A^1$ of the vector bundle will parameterize the extension 
\begin{equation}
\label{ }
\eta: \mathcal{O}_X \to \mathcal{O}_X(1)^{\oplus 5}\to T\mathbb{P}_4|_X,
\end{equation}
and the fiber will correspond to the vector space $Hom(\mathcal{O}_X(1)^{\oplus 5},\mathcal{O}_X(5))$.
It follows from the discussion above that $\eta$ allows the construction of an extension $\mathcal{W}$ as
\begin{equation}
\label{ }
\mathcal{O}_{X\times A^1} \to \mathcal{W} \to \pi^{*}_1 T\mathbb{P}_4|_X
\end{equation}
on $X \times A^1$ such that 
\begin{equation}
\label{ }
 \mathcal{W}|_{X\times 0}=\mathcal{O}_X \oplus T\mathbb{P}_4|_X,\;\;\;\mathcal{W}|_{X\times t}=(\mathcal{O}_X(1))^{\oplus 5},\,t\neq 0.
 \end{equation}
One can show that $\pi_{2*}(\mathcal{W}^{*}\otimes \pi^{*}_1\mathcal{O}_X(5))$ is a vector bundle on $A^1$ of rank $350$. The total space $W$ of this vector bundle will be the parameter space of our family. To find the universal bundle, consider more generally,  any vector bundle $\mathcal{M}$ over $\pi: X\times A \to A$, and denote the total space of $\pi_{*}\mathcal{M}$  by $M$. The fiber of $\pi_{*}\mathcal{M}$ at $a \in A$ is $H^0(X,\mathcal{M}|_{X\times a})$. Therefore we find that $p^{*}\mathcal{M}$ for 
\begin{equation}
\label{ }
{\xymatrix{    
p^{*}{\mathcal{M}} \ar[r]\ar[d]& {\mathcal{M}}\ar[d]\\
X\times M \ar[r]^{p}& X\times A
 }}
\end{equation}
has a canonical global section which  maps $(x,w)$ to $w(x)$. Returning to our original family, we find that $p^{*}(\mathcal{W}^{*}\otimes \pi^{*}_1\mathcal{O}_X(5))$ has a canonical global section, hence, we find over $X\times W$ a canonical homomorphism with kernel $\tilde{\mathcal{E}}$
\begin{equation}
\label{ }
\tilde{\mathcal{E}}\to p^{*}\mathcal{W} \to p^{*}\pi^{*}_1\mathcal{O}_X(5).
\end{equation}
It is now straightforward to see that both families $\tilde{\mathcal{F}}$ and $\tilde{\mathcal{F}}^{'}$ are contained in this family, hence its generic member will correspond to a Kodaira-Spencer class $\left(\begin{array}{c|c}0 & \alpha \\\hline \beta & *\end{array}\right).$
The generic restriction of $\tilde{\mathcal{E}}$  to $X \times w $ by  fits into the exact sequence
\begin{equation}
\label{ }
\mathcal{E} \to \mathcal{O}_{X}(1)^{\oplus 5} \to \mathcal{O}_X(5).
\end{equation}
We will use this definition of $\mathcal{E}$ to compute its cohomology.

\section{Cohomology for generic $\mathcal{E}$}\label{se-cohgen}
In this section we study the cohomology of a generic element $\mathcal{E}$ in the previously described family of 
$SU(4)$ bundles on a generic quintic $X$. It follows from our discussion in the introduction that such an 
$\mathcal{E}$ will be stable. We give an explicit parameterization of its cohomology groups in terms of  the 
bundle moduli. Various algebraic-geometric techniques are introduced  as needed. 

\subsection{$H^*(X,\mathcal{E})$}\label{cohEg}

Recall from section \ref{stableSU(4)} that the $SU(4)$ bundle is deformation of $TX \oplus \mathcal{O}_X$, given 
by the kernel of the map $\omega: \mathcal{O}_X^{\oplus 5}(1) \to \mathcal{O}_X(5)$. Note that $w$ is given by 
five global sections $s_i,\;i=1,\ldots,5$ in $H^0(X,\mathcal{O}_X(4))$. Using the  definitions
\begin{equation}
\label{ }
P=\mathcal{O}_X^{\oplus 5}(1),\;\;\;\;V=\mathcal{O}_X(5)
\end{equation}
we find $\mathcal{E}$ fits in the exact sequence
\begin{equation}
\label{defE}
{\xymatrix{
 {\mathcal{E}} \ar[r] & P \ar[r]^w & V. }}
 \end{equation}
Using upper-semi continuity the dimension of the various cohomology groups of a generic $\mathcal{E}$ must be smaller or equal  than the dimensions of the cohomology groups of $TX \oplus \mathcal{O}_X$. For convenience we recall
\begin{equation}
\label{101}
h^0(X,TX)=0,\;h^1(X,TX)=101,\;h^2(X,TX)=1,\;h^3(X,TX)=0.
\end{equation}
and
\begin{equation}
\label{hiOhiTX }
h^0(X,\mathcal{O}_X)=1,\;h^1(X,\mathcal{O}_X)=0,\;h^2(X, \mathcal{O}_X)=0,\;h^3(X,\mathcal{O}_X)=1.
\end{equation}
Observe  that
\begin{equation}
\label{ }
H^i(X,\mathcal{O}_X(n))=0,\;\;i>0, n \ge 0  \;\;\ \text{or}  \;\;\ i<3, n \le 0.
\end{equation}
This is clear for $n=0$, and for non-zero $n$, it follows simply from Kodaira vanishing theorem and Serre duality.
In particular, we find
\begin{equation}
\label{vanishcohPaV }
H^i(X,P)=0,\;\;H^i(X,V)=0,\;\;i>0.
\end{equation}
Hence the long exact sequence in cohomology related to (\ref{defE}) reduces to 
\begin{equation}
\label{ }
{\xymatrix{    
0 \ar[r]&  {H^0(X,\mathcal{E})} \ar[r] & {H^0(X,P)} \ar[r] & {H^0(X,V)}  \ar `r/10pt[d] `[l]  `^dl[dlll]  `^dr/14pt[dll]   [dll] \\ 
&  {H^1(X,\mathcal{E})} \ar[r]   & 0 &  & & }}
\end{equation}
The only non-vanishing cohomology group of $\mathcal{E}$ is $H^1(X,\mathcal{E})$, given as a quotient of
\begin{equation}
\label{P/V}
{\xymatrix{
 {H^0(X,P)} \ar[r]^{w} & {H^0(X,V)} \ar[r] & {H^1(X,\mathcal{E}}) \ar[r] & 0 , }}
\end{equation}
where the map $w$ was given in the defining equation above and represents the explicit dependence on the  vector bundle moduli. 
To parameterize ${H^0(X,P)}$ and ${H^0(X,V)}$ we will use the exact sequence
\begin{equation}
\label{defQ }
{\xymatrix{
{\mathcal{O}}_{\mathbb{P}^4}(p-5) \ar[r]^X & {\mathcal{O}}_{\mathbb{P}^4}(p) \ar[r] & {\mathcal{O}}_X(p). }}
\end{equation}
on $\mathbb{P}^4$ given by the multiplication with the defining equation $X$ of the quintic. Using its induced long exact sequence on cohomology we find
\begin{equation}
\label{ }
{\xymatrix{
H^0({\mathbb{P}}^4,{\mathcal{O}}_{\mathbb{P}^4}) \ar[r]^X &H^0({\mathbb{P}^4},{\mathcal{O}}_{\mathbb{P}^4}(5)) \ar[r] & H^0(X,V) \ar[r] & 0. }}
\end{equation}
and
\begin{equation}
\label{ }
H^0({\mathbb{P}^4},{\mathcal{O}}_{\mathbb{P}^4}(1)^{\oplus 5})  = H^0(X,P). 
\end{equation}
Also, it follows from this argument that the map $w$ is a restriction from a map $\omega_{\mathbb{P}^4}$ on $\mathbb{P}^4$.
We are left to describe $H^0(\mathbb{P}^4,\mathcal{O}_{\mathbb{P}^4}(m))$. More generally, consider the polynomial ring 
\begin{equation}
\label{ }
\mathbb{C}[x_0,\ldots,x_l],
\end{equation}
generated by the coordinates of $\mathbb{P}^l$. The global sections of $\mathcal{O}_{\mathbb{P}^l}(m)$ are simply the homogeneous elements of this ring of degree $m$. In particular,
\begin{equation}
\label{ }
h^0(\mathbb{P}^l,\mathcal{O}_{\mathbb{P}^l}(m))=\left(\begin{array}{c}l+m \\m\end{array}\right).
\end{equation}
Therefore we find 
\begin{equation}
\label{ }
{h^0(X,P)}=25,\;\;\; {h^0(X,V)}=125,
\end{equation}
hence,
\begin{equation}
\label{ }
h^1(X,\mathcal{E})=100.
\end{equation}
Its explicit parameterization is given as a quotient of the vector space of all homogenous polynomials in the 
coordinates of $\mathbb{P}^4$ of degree five (modulo the unique relation imposed by the quintic) modulo five 
sets of all homogenous linear polynomials in the coordinates of $\mathbb{P}^4$ with the map explicitly given 
by $w$. Comparing the dimensions of the cohomology groups of $\mathcal{E}$ with \ref{101} and \ref{hiOhiTX } gives an 
answer consistent with semicontinuity and preservation of Euler characteristic.

\subsection{$H^*(X,\wedge^2\mathcal{E})$}

To compute the cohomology of higher antisymmetric powers of $\mathcal{E}$ we observe that the map (\ref{defE}) 
induces the exact sequence
\begin{equation}
\label{defw2E }
{\xymatrix{    
\wedge^2{\mathcal{E}}   \ar[r] & \wedge^2 P \ar[r] & {\mathcal{E}}\otimes V. \\
}}
\end{equation}
In order to study the cohomology of $\wedge^2{\mathcal{E}} $, let us first consider the cohomology of $\wedge^2 P $. Since $\wedge^2 P \approx \mathcal{O}_X(2)^{\oplus 10}$ it follows that \begin{equation}
\label{ }
h^0(X,\wedge^2 P)=150,\;\;h^i(X, \wedge^2 P)=0,\;\;i>0.
\end{equation}
Hence the long exact sequence with respect to (\ref{defw2E }) implies
\begin{equation}
\label{cohw2E }
{\xymatrix{    
0 \ar[r]&  {H^0(X,\wedge^2 \mathcal{E})} \ar[r] & {H^0(X,\wedge^2 P)} \ar[r] & {H^0(X,\mathcal{E}\otimes V)}  \ar `r/10pt[d] `[l]  `^dl[dlll]  `^dr/14pt[dll]   [dll] \\ 
&  {H^1(X,\wedge^2 \mathcal{E})} \ar[r] & 0 \ar[r] & {H^1(X,\mathcal{E}\otimes V)}  \ar `r/10pt[d] `[l]  `^dl[dlll]  `^dr/14pt[dll]   [dll] \\ 
&  {H^2(X,\wedge^2 \mathcal{E})} \ar[r] & 0, &&&  }}
\end{equation}
Since $\wedge^2 \mathcal{E}$ is stable, $H^0(X,\wedge^2\mathcal{E})$ vanishes and
we are left to study the cohomology of ${\mathcal{E}}\otimes V$.
We consider
\begin{equation}
\label{defEV}
{\xymatrix{
 {\mathcal{E}}\otimes V \ar[r] & P\otimes V \ar[r]^{w} & V\otimes V. }}
 \end{equation}
and its corresponding long exact sequence in cohomology
\begin{equation}
\label{cohEV }
{\xymatrix{    
0 \ar[r]&  {H^0(X,\mathcal{E}\otimes V)} \ar[r] & {H^0(X,P\otimes V)} \ar[r]^{w} & {H^0(X,V\otimes V)}  \ar `r/10pt[d] `[l]  `^dl[dlll]  `^dr/14pt[dll]   [dll] \\ 
&  {H^1(X,\mathcal{E}\otimes V)} \ar[r]   & 0. &  & & }}
\end{equation}
The map $w: {H^0(X,P\otimes V)} \to {H^0(X,V\otimes V)} $ maps a $1025$ dimensional vector space to a $875$ dimensional vector space. We will show in the next subsection that for a generic quintic $X$ and a generic $w$, this map is surjective. This implies, that all cohomology of $\wedge^2\mathcal{E}$ vanishes.
\begin{equation}
\label{ }
H^i(X,\wedge^2\mathcal{E})=0,\;\forall i.
\end{equation}

\subsubsection{${w: H^0(X,P\otimes V)} \to {H^0(X,V\otimes V)} $}\label{wH0PVtoH0VV}
In this subsection we will  study the map ${w: H^0(X,P\otimes V)} \to {H^0(X,V\otimes V)} $. We will see that, 
for generic $X$ and generic $w$, it is surjective. By the openness of this condition, we have to find one specific 
$X$ with a specific $\mathcal{E}$ such that $w$ is surjective.
To begin with, note that
\begin{equation}
\label{ }
P\otimes V=\mathcal{O}_X(6)^{\oplus 5},\;\;\; V\otimes V = \mathcal{O}_X(10)
\end{equation}
We will study $w$ via pull-back to  $\mathbb{P}^4$. Using (\ref{defQ }) we find the resolutions of 
$H^0(X,P\otimes V)$ and $H^0(X,V\otimes V)$ in the commuting diagram
\begin{equation}
\label{ }
{\xymatrix{    
 H^0(X,P\otimes V) \ar[r]^{w} &  H^0(X,V\otimes V) \\
 H^0({\mathbb{P}}^4,{\mathcal{O}}_{\mathbb{P}^4}(6)^{\oplus 5}) \ar[r]^{w_{\mathbb{P}^4}}\ar[u]& \ar[u] H^0({\mathbb{P}}^4,{\mathcal{O}}_{\mathbb{P}^4}(10)) \\
H^0({\mathbb{P}}^4,{\mathcal{O}}_{\mathbb{P}^4}(1)^{\oplus 5})\ar[r] \ar[u]^X& H^0({\mathbb{P}}^4,{\mathcal{O}}_{\mathbb{P}^4}(5)), \ar[u]^X
}}
\end{equation}
where all vertical lines are short exact. To study $w_{\mathbb{P}^4}$  consider, more generally, the 
homogeneous coordinate ring
\begin{equation}
\label{ }
S := \mathbb{C}[x_0,\ldots,x_l]
\end{equation}
of $\mathbb{P}^l$. Consider a  homogenous ideal $J \subset S$ generated by a regular sequence ${f_1,\ldots,f_n}$ of $n$ homogeneous polynomials \cite{GH}. For example, the derivatives of a smooth hypersurface form a regular sequence and its corresponding ideal is called Jacobian ideal of the hypersurface. If the ideal $J$ is generated by a regular sequence, we have the Koszul complex for $R=S/ I$ \cite{GH}, that is
\begin{equation}
\label{ }
S\otimes \wedge^n E \to \ldots \to S\otimes \wedge^2 E \to S\otimes E \to  S \to R,
\end{equation}
where all maps are exact.
$E$ is simply an $n$ dimensional vector space with basis $\{e_i\}$ and the map from $S\otimes E \to  S$ is  given by $e_i \to f_i$. Note that the Koszul resolution preserves the grading of $S$. 

Returning to the case of the quintic in $\mathbb{P}^4$, let as assume  that the five global sections $s_i$ defining $w$ are derivatives of the Fermat   quintic. We find 
\begin{equation}
\label{basisdiagramm }
{\xymatrix{    
 & H^0(X,P\otimes V) \ar[r]^{w} &  H^0(X,V\otimes V)& \\
H^0({\mathbb{P}}^4,{\mathcal{O}}_{\mathbb{P}^4}(2))\otimes \wedge^2 E \ar[r] & H^0({\mathbb{P}}^4,{\mathcal{O}}_{\mathbb{P}^4}(6))\otimes E \ar[r]^{w_{\mathbb{P}^4}}\ar[u] & H^0({\mathbb{P}}^4,{\mathcal{O}}_{\mathbb{P}^4}(10)) \ar[u]^r\\
& H^0({\mathbb{P}}^4,{\mathcal{O}}_{\mathbb{P}^4}(1))\otimes E \ar[r] \ar[u]& H^0({\mathbb{P}}^4,{\mathcal{O}}_{\mathbb{P}^4}(5)). \ar[u]^X &
}}
\end{equation}
In terms of dimensions, this diagram reads as
\begin{equation}
\label{ }
{\xymatrix{    
 & H^0(X,P\otimes V) \ar[r]^{w} &  H^0(X,P\otimes V)& \\
150\ar[r] & 1050 \ar[r]^{w_{\mathbb{P}^4}}\ar[u] & 1001\ar[u]^r\\
& 25 \ar[r] \ar[u]& 126\ar[u]^X &
}}
\end{equation}
In order to prove surjectivity of $w$ we are left to find an $X$ such that  the map
\begin{equation}
\label{ }
w_{\mathbb{P}^4}\oplus X: H^0({\mathbb{P}}^4,{\mathcal{O}}_{\mathbb{P}^4}(6))\otimes E \oplus H^0({\mathbb{P}}^4,{\mathcal{O}}_{\mathbb{P}^4}(5)) \to H^0({\mathbb{P}}^4,{\mathcal{O}}_{\mathbb{P}^4}(10))
\end{equation}
is surjective.
The ideal spanned by $w_{\mathbb{P}^4}$ in $H^0({\mathbb{P}}^4,{\mathcal{O}}_{\mathbb{P}^4}(10))$ is of the form $\sum_i (x_i)^4$. Hence we are missing all monomials in $\{x_i\}_{i}$ of degree ten whose highest coefficient is smaller than four. This is a $101$ dimensional vector space and a convenient parameterization (up to permutation) is given by 
\begin{equation}
\label{ }
M=\sum_{i=1}^5 M_i=(33310)\oplus (33220)\oplus (33211)\oplus (32221)\oplus (22222)
\end{equation}
where $i_0...i_5$ stands for $\prod_{j=0}^5 x_j^{i_j}$ and $(i_0...i_5)$ stands for all permutations.
Similarly we can describe all monomials in $H^0({\mathbb{P}}^4,{\mathcal{O}}_{\mathbb{P}^4}(5))$ which are not in the the ideal generated by $w$. They are simply all monomials in $\{x_i\}_{i}$ of degree five whose highest coefficient is smaller than four. Using the same convention as above, a convenient parameterization is given by
\begin{equation}
\label{ }
N=\sum_{i=1}^5 N_i=(32000)\oplus(31100)\oplus(22100)\oplus(21110)\oplus(11111)
\end{equation}
Both $N$ and $M$ are $101$ dimensional vector spaces and we have to find a $X: N \to M$ which is an isomorphism. Lets chose $X=\prod x_i$. The induced map splits in the direct sum 
\begin{equation}
\label{ }
X=\oplus_i X_i : \sum_{i} N_i \to \sum_{i} M_i
\end{equation}
where $X_1$ and $X_2$ are the zero map and $X_i,\;i=3,4,5$ are isomorphisms. Being invertible on $N_i,\;i=3,4,5$ is an open condition, hence for a sufficiently small deformation of $X$ we keep this property. Explicitly we deform $X$ by
\begin{equation}
\label{ }
X=\prod x_i + \sum_{i\leqslant j\;k }\epsilon_{ij,k}v_{ij,k}
\end{equation}
where $v_{ij,k}$ stands for the monomial $x_ix_jx_k^3$ and $\epsilon_{ij,k}$ are sufficiently small parameters. The induced map is
again a direct sum $X=\oplus_i X_i$. We will show that is induced map on $N_1$ and $N_2$ is also an isomorphism.
It is not to difficult to show that the $30 \times 30$ matrix $X_2$ splits into $10$ blocks of $3 \times 3$ matrizes. For example for the three monomials $\{31100, 01103, 01130\}\in N_2 $ we find as the only non-vanishing components of $X_2$ 
\begin{equation}
\label{ }
\left(\begin{array}{c|c|c|c}  & 02203 & 32230 & 32203 \\\hline 31100 & 0 & \epsilon_{23,4} & \epsilon_{23,5} \\\hline 01103 & \epsilon_{23,4} & 0 & 0 \\\hline 01130 & \epsilon_{23,5} & \epsilon_{23,1} & 0\end{array}\right).
\end{equation}
Note that the determinant of this $3 \times 3$ matrix is $\epsilon_{23,4}\epsilon_{23,5} \epsilon_{23,1}$. It 
follows from this discussion that $X_2$ will be an isomorphism iff $\epsilon_{ij,k}\neq 0,\;\forall i,j,k$.
Unfortunatly, the $20 \times 20$ matrix $X_1$ does not seem to have such a clear structure. Nevertheless, a quick 
check in Maple ensures that $X_1$ is an isomorphism as well. 

\section{A non-generic $E$ on the Dwork quintic}\label{sect-Es}
In the previous section we have shown that $H^*(X,\wedge^2 \mathcal{E})$ vanishes for generic $\mathcal{E}$ on 
generic quintics $X$. In this section we show that this is only a generic result. We give a concrete example of 
$\mathcal{ E}$ on the Dwork quintic threefold, such that
\begin{equation}
\label{ }
h^1(X,\wedge^2 \mathcal{E})=h^2(X,\wedge^2 \mathcal{E})=50.
\end{equation}
Since this $\mathcal{E}$ is not generic, we will include a proof of its stability. We will see later that the 
Yukawa coupling for this vector bundle also does not vanish.

Our bundle is given by
\begin{equation}
\label{defEs}
{\xymatrix{
 {\mathcal{E}} \ar[r] & P \ar[r]^w & V, }}
 \end{equation}
where $w$ is given by the Jacobian ideal of the Fermat quintic. We will consider this bundle on the Dwork 
quintic, which is given by the polynomial
\begin{equation}
\label{ Dwork}
\sum x_i^5+\prod x_i=0.
\end{equation}
To begin with let us prove the stability of $\mathcal{E}$. It follows from the appendix that it is sufficient to show that the maps
\begin{equation}
\label{condstab1}
H^0(X,P) \to H^0(X,V)
\end{equation}
and 
\begin{equation}
\label{condstab2}
H^0(X,\wedge^2 P) \to H^0(X,P\otimes V) 
\end{equation}
are injective.
To check these conditions we will use techniques developed in section~\ref{wH0PVtoH0VV}. In particular, to show 
(\ref{condstab1}), we use the resolution
\begin{equation}
\label{condstab1diagram }
{\xymatrix{    
 & H^0(X,P) \ar[r]^{w} &  H^0(X, V)& \\
& H^0({\mathbb{P}}^4,{\mathcal{O}}_{\mathbb{P}^4}(1))\otimes E \ar[r]^{w_{\mathbb{P}^4}}\ar[u] & H^0({\mathbb{P}}^4,{\mathcal{O}}_{\mathbb{P}^4}(5)) \ar[u]^r\\
& & H^0({\mathbb{P}}^4,{\mathcal{O}}_{\mathbb{P}^4}). \ar[u]^X &
}}
\end{equation}
where $w_{\mathbb{P}^4}$ is given by the five partial derivatives of the Fermat quintic and $X$ by the Dwork 
(\ref{ Dwork}). To prove injectivity of $w$ we have to show that $X$ is not in the ideal generated by 
$w_{\mathbb{P}^4}$. But what is in the Jacobian ideal of the Fermat quintic? Certainly the Fermat quintic 
itself. Borrowing a result from \cite{D} shows that a quintic will be in this ideal iff it is isomorphic to the 
Fermat quintic. Since the Dwork quintic is not isomorphic to the Fermat, the image of $w_{\mathbb{P}^4}$ does not 
intersect the image of $X$, hence $w$ is injective.
To show that condition (\ref{condstab2}) is obeyed we will show that the map 
\begin{equation}
\label{}
H^0(X,P\otimes P) \to H^0(X,P\otimes V) 
\end{equation}
is injective. To guarantee this we simply need to show the injectivity of the map
\begin{equation}
\label{}
H^0(X, P(1)) \to H^0(X, V(1)).
\end{equation}
Again, pulling it back to $\mathbb{P}_4$ we find
\begin{equation}
\label{ }
{\xymatrix{    
 & H^0(X,P(1)) \ar[r]^{w} &  H^0(X, V(1))& \\
& H^0({\mathbb{P}}^4,{\mathcal{O}}_{\mathbb{P}^4}(2))\otimes E \ar[r]^{w_{\mathbb{P}^4}}\ar[u] & H^0({\mathbb{P}}^4,{\mathcal{O}}_{\mathbb{P}^4}(6)) \ar[u]^r\\
& & H^0({\mathbb{P}}^4,{\mathcal{O}}_{\mathbb{P}^4}(1)). \ar[u]^X &
}}
\end{equation}
where $w_{\mathbb{P}^4}$  and $X$ are as above. Recall from section~\ref{wH0PVtoH0VV} that the image of $w_{\mathbb{P}^4}$ misses all monomials of degree six whose highest power is smaller then 4. The image of $X$ in that quotient is given by degree six polynomials whose highest power is two. Hence these images don't intersect and $w$ is injective.

What about the cohomology of $\mathcal{E}$? It is easy to see that all arguments of section \ref{cohEg} apply. We recall the results for convenience. All cohomology of $\mathcal{E}$ vanishes except $H^1(X,\mathcal{E})$ which can be explicitly parameterized as the quotient of 
\begin{equation}
\label{ }
{\xymatrix{
 {H^0(X,P)} \ar[r]^{w} & {H^0(X,V)} \ar[r] & {H^1(X,\mathcal{E}}).}}
\end{equation}
Let us consider the cohomology of $\wedge^2 \mathcal{E}$. Tracing through (\ref{cohw2E }) and (\ref{cohEV }) we find that
\begin{equation}
\label{ }
H^1(X,\wedge^2 \mathcal{E})=\frac{ker(H^0(X,P\otimes V)\to H^0(X,V\otimes V))}{H^0(X,\wedge^2 P)}
\end{equation}
and
\begin{equation}
\label{h2w2E }
H^2(X,\wedge^2 \mathcal{E})=\frac{H^0(X,V\otimes V)}{H^0(X,P\otimes V)}.
\end{equation}
To determine the dimensionality we consider the diagram (\ref{basisdiagramm }) and the remarks which followed. We find
\begin{equation}
\label{ }
h^1(X,\wedge^2 \mathcal{E})=h^2(X,\wedge^2 \mathcal{E})=50.
\end{equation}

\section{Yukawa couplings}\label{se-Yukawa}

In this section we make some comments about the Yukawa couplings corresponding to the triple product
\begin{equation}
\label{ triple}
H^1(X,\mathcal{E})\otimes H^1(X,\mathcal{E})\otimes H^1(X,\wedge^2\mathcal{E}) \to H^3(X,\wedge^4\mathcal{E})\cong \mathbb{C}.
\end{equation}
In particular we give their  explicit dependence on the vector bundle moduli, that is, their dependence on
\begin{equation}
\label{ }
w \in H^0(X,\mathcal{O}_X(4)^{\oplus 5}).
\end{equation}
First note that the triple product (\ref{ triple}) can be rewritten as the pairing:
\begin{equation}
\label{ }
H^1(X,\mathcal{E})\otimes H^1(X,\mathcal{E}) \to H^2(X,\wedge^2\mathcal{E}).
\end{equation}
By (\ref{P/V}) and (\ref{h2w2E }), this becomes a pairing:
\begin{equation}
\label{ }
\frac{H^0(X,V)}{H^0(X,P)}\otimes \frac{H^0(X,V)}{H^0(X,P)} \to 
\frac{H^0(X,V\otimes V)}{H^0(X,V\otimes P)}.
\end{equation}
By tracing through a few commuting diagrams, we identify this pairing as the natural multiplication. 
It follows from (\ref{P/V}) and (\ref{h2w2E }) that the inclusions 
of the denominators into the numerators are given by the map $w$ defining $\mathcal{E}$. In particular, 
the dependence of the Yukawa couplings on the moduli space of $\mathcal{E}$ is made manifest.

A well known special case occurs for the specific choice of $w$ which makes $\mathcal{E}$ an extension 
of the tangent bundle by the trivial bundle. Recall from (\ref{Fasext}) that this $\mathcal{E}$ fits 
into the commuting diagram
\begin{equation}
\label{ }
{\xymatrix{    
{\mathcal{O}}_X \ar[r]\ar[d] &{\mathcal{O}}_X \ar[d]& \\
{\mathcal{E}}  \ar[r]\ar[d] &{\mathcal{O}}_X(1)^{\oplus 5} \ar[r]\ar[d]^{\nabla_i}
&{\mathcal{O}}_X(5)\ar[d] \\
TX \ar[r] &T{\mathbb{P}_4}|_X \ar[r]^{\nabla_i Q} &{\mathcal{O}}_X(5).
}}
\end{equation}
In particular, it follows from this that $w$ is given by the partials of the quintic $X$. It follows from the 
long exact sequence in cohomology corresponding to the first vertical short exact sequence that\begin{equation}
\label{ }
H^1(X,\mathcal{E}) = H^1(X,TX),\;\;\;H^2(X,\wedge^2\mathcal{E}) = H^2(X,\wedge^2 TX)
\end{equation}
Hence for this particular $\mathcal{E}$ the pairing is given by
\begin{equation}
\label{pairTX }
H^1(X,TX)\otimes H^1(X,TX) \to H^2(X,\wedge^2 TX).
\end{equation}
But now we are in the situation studied by Carlson and Griffiths, \cite{CG,G}. Their result states that the pairing 
{pairTX } is given by the polynomial multiplication of the rings
\begin{equation}
\label{ }
\frac{H^0(X,\mathcal{O}_X(5))}{H^0(X,\mathcal{O}_X(1)^{\oplus 5})}\otimes \frac{H^0(X,\mathcal{O}_X(5))}{H^0(X,\mathcal{O}_X(1)^{\oplus 5}} \to 
\frac{H^0(X,\mathcal{O}_X(10))}{H^0(X,\mathcal{O}_X(6)^{\oplus 5})},
\end{equation}
where the maps $H^0(X,\mathcal{O}_X(1)^{\oplus 5}) \to H^0(X,\mathcal{O}_X(5))$ and  
$H^0(X,\mathcal{O}_X(6)^{\oplus 5}) \to H^0(X,\mathcal{O}_X(10))$ are given by partials of $X$.

\section{Non-vanishing Yukawa couplings for all quintics $X$}\label{se-genquintic}

We now generalize the result of the previous section: we show the existence of a stable bundle $\mathcal{E}$ 
with non-vanishing $H^2(X,\mathcal{E})$ for every smooth quintic. In particular, this implies  the existence 
of heterotic vacua with non-vanishing Yukawa coupling for every quintic.
 
To do so we use the analytic result by Li and Yau \cite{LY} stated in the introduction.  Recall that the sufficient 
condition for the existence of a hermitian Yang-Mills connection and a solution for Strominger's equation on 
$\mathcal{E}$ was that  the off diagonal terms in $Ext^1(TX\oplus \mathcal{O}_X,TX\oplus \mathcal{O}_X)$  do not 
vanish. A sufficient condition for the non-vanishing of the anti-diagonal is that neither $\mathcal{E}$ nor 
$\mathcal{E}^{*}$ admits global sections. The vanishing of $h^0(X,\mathcal{E}^{*})=h^3(X,\mathcal{E})$  follows 
from its definition   (\ref{defE}) and the vanishing theorem (\ref{vanishcohPaV }). The condition for the vanishing 
of global sections of $\mathcal{E}$ was analyzed in \ref{condstab1}. In particular, in diagram (\ref{condstab1diagram })  
we choose the map $X$ to be the defining equation of our quintic. For the map $w_{\mathbb{P}_4}$ we use the partials of 
a different quintic $X^{'}$ which is a small deformation of $X$, that is 
\begin{equation}
\label{ }
X{'}=X+\epsilon
\end{equation}
If we choose a smooth quintic $X^{'}$ which is not isomorphic to $X$, it follows from the arguments 
presented in section~\ref{sect-Es} that $\mathcal{E}$ has no global sections. Note that the set of all 
quintics $X^{'}$ is a $126$ dimensional space, and teh condition is violated on a $25$ dimensional subset.
Note that it follow from this discussion that our example in section~\ref{sect-Es} can be made to fit in 
this framework.

To show the non-vanishing of $H^2(X,\wedge^2 \mathcal{E})$ we recall its description in (\ref{h2w2E }). To 
compute it we have to trace through diagram (\ref{basisdiagramm }). We have to show that
\begin{equation}
\label{ }
im\;\left( w_{\mathbb{P}_4} : H^0(\mathbb{P}_4,\mathcal{O}_{\mathbb{P}_4}(5)\right)\to H^0(\mathbb{P}_4,\mathcal{O}_{\mathbb{P}_4}(10)) \bigcup X \cdot H^0(\mathbb{P}_4,\mathcal{O}_{\mathbb{P}^4}(5))
\end{equation}
does not span the $1001$ dimensional vector space $H^0(\mathbb{P}_4,\mathcal{O}_{\mathbb{P}^4}(10))$ for some 
$X^{'}$. Note that the image of $w_{\mathbb{P}^4}$ is $900$ dimensional and the image of the fixed space 
$X \cdot H^0(\mathbb{P}_4,\mathcal{O}_{\mathbb{P}^4}(5))$ is $126$ dimensional. Therefore we have to study the 
sublocus in the Grassmannian $G(900,1001)$ of subspaces which do not intersect the $126$ dimensional vector space 
transversally. We find its codimension to be $26$. Among all quintics $X^{'}$, which depend on $126$ parameters, 
each component of those not transversal to the $126$ dimensional space is at least  $100$ dimensional. (Since 
$X$ is contained in this locus, it is not empty). Therefore we can find $X^{'}$ such that $H^0(X,\mathcal{E})=0$ 
and $H^1(X,\wedge^2\mathcal{E})\neq0$.

\vskip 0.2in
\noindent {\bf Acknowledgements}

\noindent
We thank M. Douglas and E. Sharpe for valuable discussions. 
R.D. is partially supported by NSF grants DMS 0104354 and DMS 0612992, and
by NSF Focused Research Grant DMS 0139799 "The Geometry of Superstrings".
The work of R.R. was supported in part by DOE grant DE-FG02-96ER40959.
In addition, R.R. would like to thank the hospitality of Harvard math department where part of this work was done.

\vskip 0.2in
\noindent {\bf Appendix}

In this section we will consider the stability of $\mathcal{E}$ on the quintic $X$. To begin with, we will show that the conditions
\begin{equation}
\label{stabilcond}
0=H^0(X,\mathcal{E})=H^0(X,\wedge^2\mathcal{E})=H^0(X,\wedge^3\mathcal{E})
\end{equation}
are sufficient to guarantee stability of $\mathcal{E}$.  
More generally, to define Mumford-Takemoto stability on a torsion free coherent sheaf $W$ on a complex Kaehler manifold of dimension $n$ one introduces its slope with respect to a Kaehler class $H$
\begin{equation}
\label{ }
\mu_H(W)=\frac{c_1(W)\cdot H^{n-1}}{\rk(W)}.
\end{equation}
A sheaf  is stable if  all torsion free coherent sub-sheaves $\mathcal{F}$ whose rank is smaller the the rank of $W$ obey 
\begin{equation}
\label{ }
\mu_H(\mathcal{F})<\mu_H(W).
\end{equation}
In general the condition of stability will depend on the chosen Kaehler class $H$. However, if we restrict $X$ to be 
a smooth quintic in $\mathbb{P}_4$, then the Lefschetz Hyperplane theorem ensures that $Pic(X)=\mathbb{Z}$ and 
$h^{1,1}(X)=\mathbb{C}$. Therefore, (up to re-scaling), $H$ is uniquely determined and, henceforth, we will 
suppress it. 

Assume there exists a torsion free destabilizing sub-sheaf $\mathcal{F}$ of rank $r$ of $\mathcal{E}$,  that is
\begin{equation}
\label{ }
\mathcal{F} \subset \mathcal{E}
\end{equation} 
with $\mu(\mathcal{F}) \ge 0$. Therefore there is the non-zero map
\begin{equation}
\label{ }
\wedge^r \mathcal{F} \to \wedge^r \mathcal{E}
\end{equation}
and a non-zero map
\begin{equation}
\label{ }
(\wedge^r \mathcal{F})^{**} \to \wedge^r \mathcal{E}^{**}=\wedge^r \mathcal{E}
\end{equation}
Note that for torsion free sheaves the  first Chern class of $\mathcal{F}$ can be defined \cite{K} by 
\begin{equation}
\label{ }
c_1(\mathcal{F})=c_1((\wedge^r \mathcal{F})^{**}).
\end{equation}
Since $(\wedge^r \mathcal{F})^{**}$ is a reflexive torsion free sheaf of rank one, it is a line bundle. Therefore we find a destabilizing line bundle 
\begin{equation}
\label{ }
(\wedge^r \mathcal{F})^{**} \subset \wedge^r \mathcal{E}.
\end{equation}
It follows from this discussion that in order to prove stability of $\mathcal{E}$, we simply have to show that no 
destabilizing line bundle of $\wedge^i \mathcal{E},\;i=1,2,3$ exists. Recall from above that every line bundle on 
$X$ is of the form $\mathcal{O}_X(n)$ with the property that  $h^0(X,\mathcal{O}_X(n))> 0$. That is, 
every destabilizing line bundle has at least one global section. Assume that we have shown that 
$H^0(X,\wedge^i \mathcal{E})=0,\;i=1,2,3$. Assume that there is a destabilizing line bundle
\begin{equation}
\label{ }
\mathcal{O}_X(n) \subset \wedge^r \mathcal{E}
\end{equation}
for some $0<r<4$. Then we have the inclusion
\begin{equation}
\label{ }
H^0(X,\mathcal{O}_X(n)) \subset H^0(X,\mathcal{E}),
\end{equation}
a contradiction. Hence, (\ref{stabilcond}) is sufficient to guarantee the stability of $\mathcal{E}$.

To show that $H^0(X,\wedge^i \mathcal{E})=0,\;i=1,2,3$ we recall the defining sequence (\ref{defE}). For $i=1$ we must prove that
\begin{equation}
\label{h0E }
H^0(X,P) \to H^0(X,V)
\end{equation}
is injective. It follows from (\ref{defw2E }) that in the case of  $i=2$ we must show that
\begin{equation}
\label{ }
H^0(X,\wedge^2 P) \to H^0(X,\mathcal{E}\otimes V)
\end{equation}
is injective. If we combine this map with the injective map
\begin{equation}
\label{ }
H^0(X,\mathcal{E}\otimes V) \to H^0(X,P \otimes V)
\end{equation}
we are left to show that the map \begin{equation}
\label{ }
H^0(X,\wedge^2 P) \to H^0(X,P\otimes V) 
\end{equation}
is injective. Note that this map is the restriction of the map
\begin{equation}
\label{h0E2}
1\otimes w : H^0(X,P\otimes P) \to H^0(X,P\otimes V).
\end{equation}
To show that $H^0(X,\wedge^3 \mathcal{E})=0,$ recall that
\begin{equation}
\label{h0E3}
H^0(X,\wedge^3 \mathcal{E})=H^0(X,\mathcal{E}^{*})=H^3(X,\mathcal{E})^{*}.
\end{equation}
The vanishing of $H^3(X,\mathcal{E})$ follows from the long exact sequence  in cohomology  associated to 
(\ref{defE}) and the vanishing of the non-zero cohomology group (\ref{vanishcohPaV }) of $P$ and $V$ .



\end{document}